# Assessing quantitative phase reliability in electron ptychography using a twisted trilayer $WS_2$ thickness staircase

Hu Bao[1,2,3], Qilong Feng[4], Kun Ye[2], Tianyu Xue[2], Stephen J. Pennycook[5], Yihan Zhu[4], Anmin Nie[2*], Shoucong Ning[1*], Jing Tao[1]

[1] Department of Physics, University of Science and Technology of China, Hefei, 230026, China

[2] Center for High Pressure Science (CHiPS), State Key Laboratory of Metastable Materials Science & Technology, Yanshan University, Qinhuangdao, 066004, China

[3] State Key Laboratory of Metastable Materials Science and Technology, Key Laboratory for Microstructural Material Physics of Hebei Province, School of Science, Yanshan University, Qinhuangdao, 066004, China

[4] College of Chemical Engineering, Zhejiang University of Technology, Hangzhou, 310014, China

[5] School of Physical Sciences and CAS Key Laboratory of Vacuum Sciences, University of Chinese Academy of Sciences, Beijing, 100049, China

**Abstract**

The phase of an electron wave is sensitive to the local electrostatic environment and can therefore provide information about atomic-scale structural variations. In electron ptychography, however, specimen thickness, multiple scattering and projected atomic overlap can complicate the quantitative interpretation of the reconstructed phase. Here, we use a twisted trilayer $WS_2$ thickness staircase containing monolayer, bilayer and trilayer regions as an internal reference, with all regions measured under identical experimental conditions. The mean W-column phase increases by approximately 13.2% from the monolayer to the bilayer region, compared with only 4.7% from the bilayer to the trilayer region, revealing a nonlinear thickness dependence. W-S overlap has little effect on the W-column phase, whereas short-range projected W-W overlap increases it by approximately 8-9%. These results provide a benchmark for evaluating quantitative phase measurement in multilayer $WS_2$ and for assessing phase-based mapping of local electrostatic potential, charge-density and three-dimensional atomic-scale polarization.

## Introduction

Material properties are governed not only by long-range crystallographic order but also by local three-dimensional structural modulations [1–4]. Variations in atomic positions, interlayer registry, lattice distortion, strain, interfaces, and defects can alter local bonding and electronic environments [4–7]. Such variations play important roles in phenomena including charge-density waves [9–12], ferroic order [13, 14], and moiré reconstruction [5, 6]. Defects and structural disorder can further perturb these ordered states or locally modify the associated structural patterns, producing variations that cannot be fully described by the mean crystal structure [9, 15–17]. Establishing the relationship between local three-dimensional structure and material response is therefore important for understanding the behavior of complex materials [1, 7, 8, 10, 18].

A range of diffraction, scattering, and phase-sensitive techniques can be used to investigate this relationship across different length scales. X-ray diffraction and scattering characterize crystal symmetry, lattice modulations and long-range order [22, 23], whereas neutron scattering probes atomic and magnetic correlations as well as lattice dynamics [24–26]. Electron phase-imaging methods complement these techniques by providing atomic-scale information about the projected electrostatic potential [20, 21, 27]. Under appropriate scattering approximations, the reconstructed electron-wave phase can be quantitatively related to the projected potential, thereby providing a means of correlating local electrostic variations with the underlying atomic structure [19, 27, 28]. In four-dimensional scanning transmission electron microscopy (4D-STEM), a two-dimensional diffraction pattern is recorded at each probe position as a focused electron beam scans the specimen [29]. Electron ptychography uses the overlap between adjacent probe positions to reconstruct the complex specimen response, including its phase [30, 31]. This approach provides a route for examining the relationship between local phase variations and three-dimensional structur [31–34]. However, the quantitative interpretation of the reconstructed phase can be complicated by specimen thickness, inerlayer propagation, multiple scattering and projected atomic overlap [28, 32, 35]. Understanding these effects is therefore essential for establishing a reliable relationship between the reconstructed phase and the local electrostatic potential [36].

Here, we use a twisted trilayer $WS_2$ (tt$WS_2$) thickness staircase as an internal reference to investigate the dependence of the reconstructed ptychographic phase on layer number and local projected atomic overlap. The specimen contains monolayer, bilayer and trilayer regions within the same field of view, enabling direct comparison under identical acquisition and reconstruction conditions. Under the weak-scattering approximation, the phase associated with the projected specimen potential is expected to increase approximately linearly with layer number. In the present specimen, however, the mean W-column phase increases by approximately 13.2% from the monolayer to the bilayer region but by only 4.7% from the bilayer to the trilayer region, revealing a nonlinear thickness dependence. Analysis of local projected configurations furhter shows that W-S overlap has little effect on the W-column phase, whereas short-range projected W-W overlap produces an enhancement of approximately 8-9%. These observations demonstrate that, multiple scattering and projected atomic overlap jointly influence the reconstructed phase in multilayer $WS_2$. The thichness staircase therefore provide an internal benchmark for evaluating quantitative phase measurements and for inerpreting phase-based mappings of local electrostatic potential, charge redistribution and three-dimensional atomic-scale polarization.

## Results and Discussion

### Ptychographic imaging of a tt$WS_2$ thickness staircase

We characterized a tt$WS_2$ thickness-staircase containing monolayer, bilayer and trilayer regions within a single field of view. Because all regions were recorded and reconstructed under identical conditions, they provide an internal reference for comparing the layer-number dependence of the ptychographic phase. In the 4D-STEM geometry, a focused electron probe scans the tt$WS_2$ specimen, while a pixelated detector records a two-dimensional diffraction pattern at each probe position (Fig. 1a). This generates a four-dimensional dataset containing both real-space scan coordinates and reciprocal-space diffraction intensities. The ADF image calculated from the 4D-STEM dataset shows the moiré superlattice and thickness-staircase regions (Fig. 1b). The ADF image iprovides an incoherent intensity projection and does not directly contain phase information associated with the local projected potential. In comparison, the ptychographic phase image provides clearer atomic-column contrast for the subsequent statistical analysis. The reconstructed phase image provides the basis for the subsequent statistical analysis of atomic-column phase values (Fig. 1c and Supplementary Sections S1 and S2). The averaged diffraction pattern exhibits vertically elongated Bragg spots associated with scan-position distortions (Fig. 1d). The Fourier transform of the reconstructed object function contains high-spatial-frequency components, indicating an extended information-transfer range and improved in-plane spatial resolution (Fig. 1e). The single-slice and multislice reconstructions produce noticeably different contrast in the thicker regions (Figs. 1f, g). The single-slice reconstruction provides high-resolution phase contrast but exhibits local anomalies, including contrast reversal, in the thicker parts of the specimen. These features are attributed to the limited ability of the single-slice model to describe interlayer propagation and multiple scattering. The multislice reconstruction reduces the contrast reversal and provides more consistent atomic-column contrast across the different thicknesses. These differences indicate that the phase contrast in multilayer regions depends on both specimen thickness and the reconstruction model.

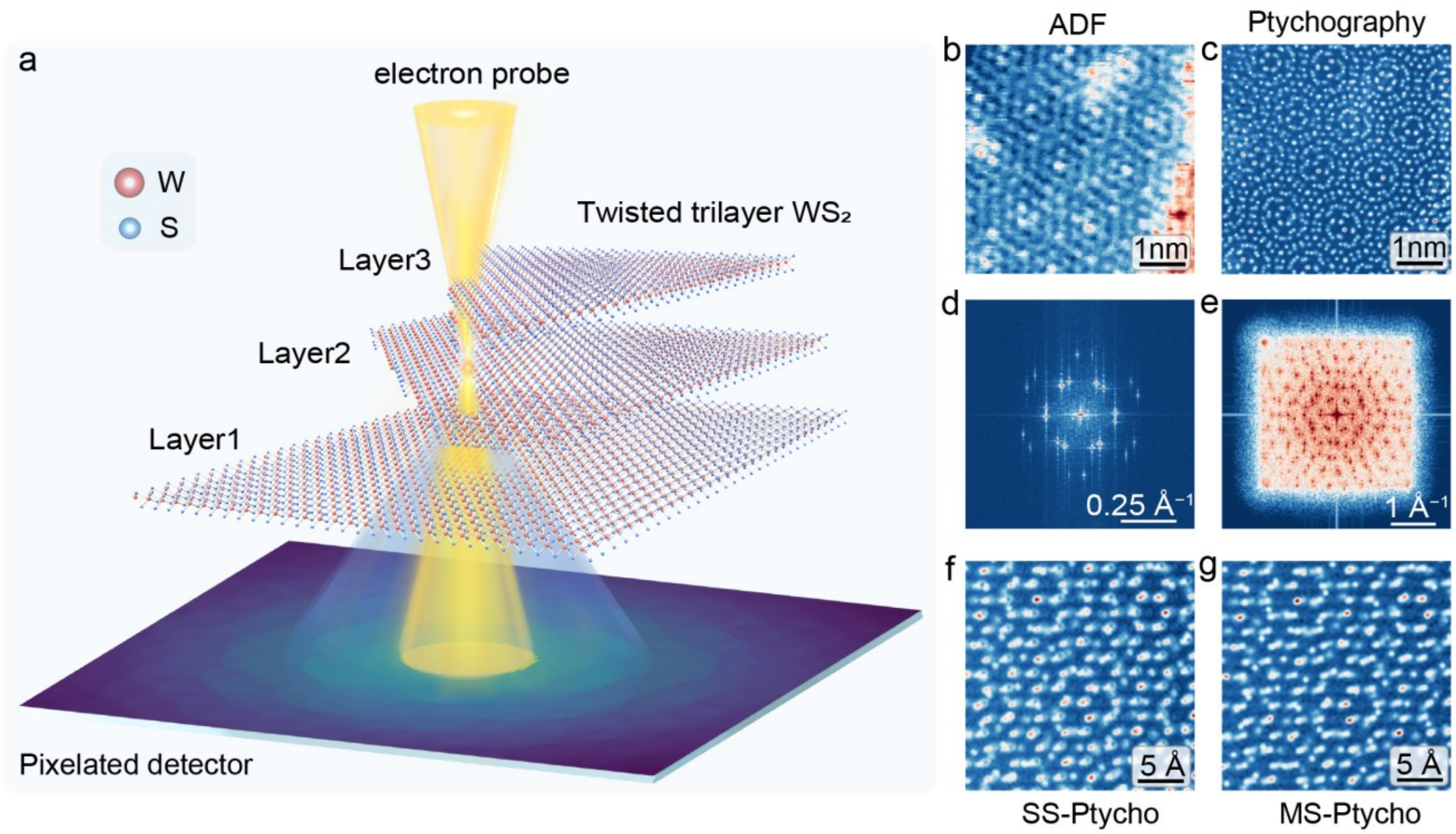

**Fig.1 | Ptychographic imaging of a ttWS$_2$ thickness staircase.** (a) Schematic of the 4D-STEM ptychography experiment. A focused electron probe scans a twisted WS$_2$ specimen containing monolayer, bilayer and trilayer regions, while a pixelated detector records a diffraction pattern at each position. (b) ADF image calculated from the 4D-STEM dataset, showing the moiré superlattice and thickness-staircase regions. (c) Reconstructed ptychographic phase image, providing atomic-column contrast for the subsequent statistical analysis. (d) Averaged 4D-STEM diffraction pattern with vertically elongated Bragg spots. (e) Fourier transform of the reconstructed object function, including high-spatial-frequency components. (f) Single-slice reconstruction, showing contrast anomalies and reversal in the thicker regions. (g) Multislice reconstruction, in which these contrast anomalies are reduced.

**Layer-number-dependent phase response in the experimental $WS_2$ thickness staircase**

We next analyzed the W-column phase values in regions of different thicknesses. Because specimen edges and local contamination can affect the extracted phase values, we first identified the boundaries of the monolayer, bilayer, and trilayer regions in the reconstructed phase image. Areas affected by specimen edges or contamination were then excluded from the analysis. (Supplementary Section S1). Figure 2a shows the monolayer, bilayer, and trilayer regions selected for analysis, outlined in red, blue, and yellow, respectively. The corresponding magnified images in Fig. 2b resolve the atomic columns used for Gaussian fitting and phase extraction (Supplementary Section S2). For comparison, the statistics obtained before excluding boundary and contaminated regions show broader distributions, with Gaussian-fitted distribution centers of approximately 0.40, 0.45 and 0.47, for the W-column phase values, respectively (Supplementary Section S3). Within the screened regions shown in Figure 2a, the W-column phase-values distribution peaks shift toward higher values from monolayer to bilayer to trilayer $WS_2$ (Fig. 2c). The corresponding mean values increases from approximately 0.38 in the monolayer region to approximately 0.43 in the bilayer region and approximately 0.45 in the trilayer region (Fig. 2d). This corresponds to an increase of about 13.2% from monolayer to bilayer, but only about 4.7% from bilayer to trilayer. Thus, although the mean W-column phase increases with layer number, the incremental increase becomes smaller at greater thickness. This nonlinear trend differs from a simple linear accumulation with layer number and is consistent with an increasing contribution from multiple scattering.

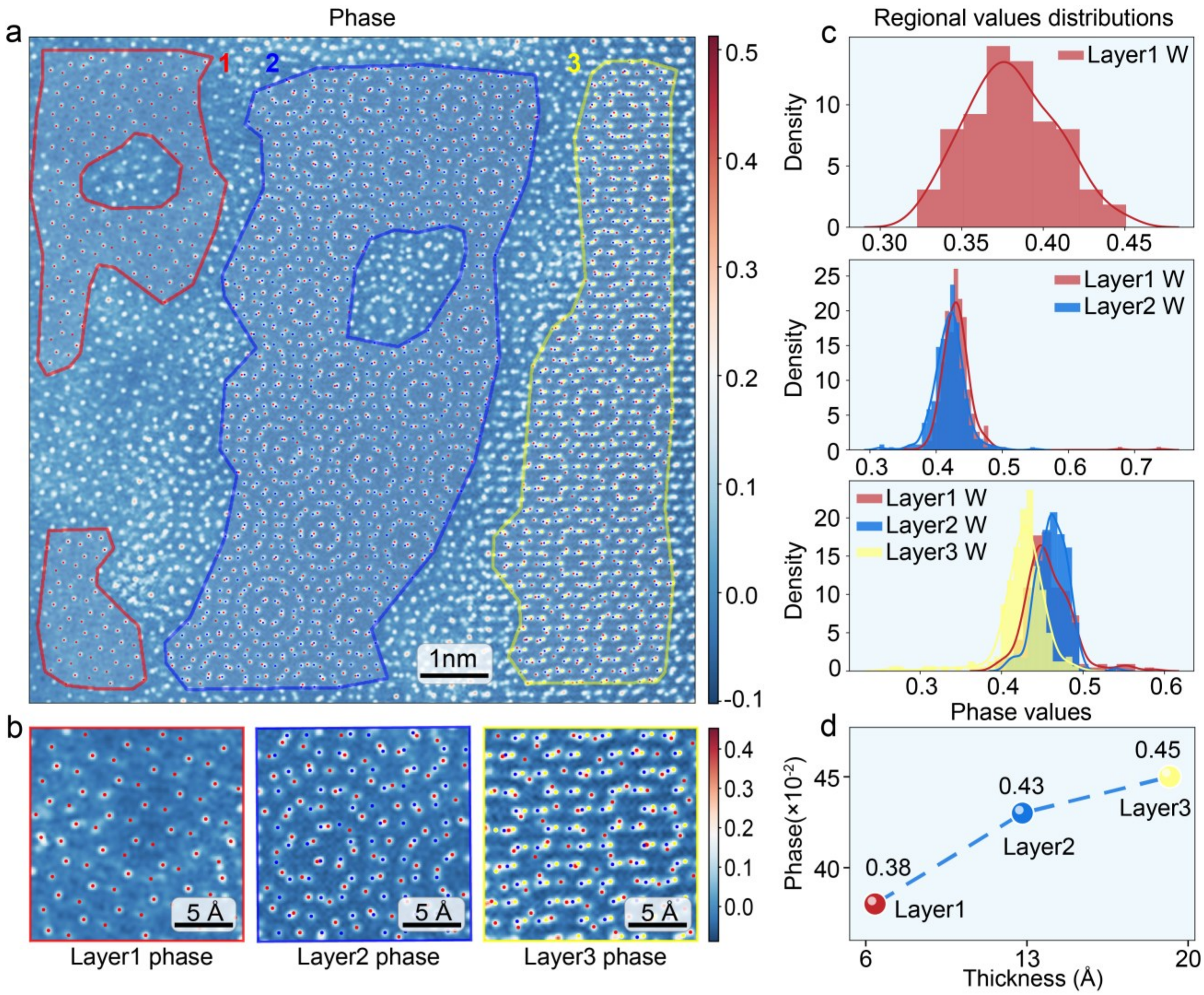

**Fig.2 | Phase-values statistics of the experimental $WS_2$ thickness-staircase regions.** (a) Ptychographic phase image of tt$WS_2$ and selected analysis regions. Red, blue and yellow outlines denote monolayer, bilayer and trilayer regions, respectively. (b) Atomic-resolution magnified images showing clear W-column localization in each region. (c) W-column phase-values distributions in the three regions, with red, blue and yellow corresponding to the monolayer, bilayer and trilayer regions, respectively. (d) Mean phase values as a function of thickness, showing an increase from approximately 0.38 in the monolayer region to 0.43 in the bilayer region and 0.45 in the trilayer region.

**W-column phase enhancement associated with local atomic overlap**

Beyond the overall thickness dependence, twisted multilayer $WS_2$ exhibits several local projected-overlap configurations. Relative rotation between layers produces several configurations along the electron-beam direction: relatively isolated W sites, projected W-S overlap and short-range projected W-W overlap. These configurations modify local scattering strength and allow us to examine the microscopic origin of phase distortions in multilayer $WS_2$. Figure 3a schematically illustrates three representative configurations: isolated W, projected W-S overlap and projected W-W overlap. To quantify the effect of projected overlap distance, we exmamined the dependence of the W-column phase on the projected W-W distance (Fig. 3b and Supplementary Section S4). The phase values exhibit a distinct variation at projected W-W distances below approximately 40 pm. Using the projected interatomic distances and local atomic identities, we classified the W sites into three configurations: isolated W, W-S overlap and W-W overlap (Supplementary Section S5). Comparing with isolated W sites reveals distinct phase responses for the W-S and W-W overlap configurations (Figs. 3c, d). In the bilayer and trilayer regions, the W-column phase associated with W-S overlap is similar to that of isolated W sites, suggesting that S atoms have only a weak influence on the W-column phase values. W-W overlap, by contrast, produces a phase enhancement. In the bilayer region, the phase values increases from approximately 0.428 for isolated W to 0.468 for W-W overlap, corresponding to an enhancement of about 9.3%. In the trilayer region, the corresponding phase values increases from approximately 0.446 to 0.482, corresponding to an enhancement of about 8.1%. These measurements demonstrate that the reconstructed W-column phase varies with both layer number and local projected overlap. Among the configurations examined, short-range W-W overlap produces the largest phase enhancement.

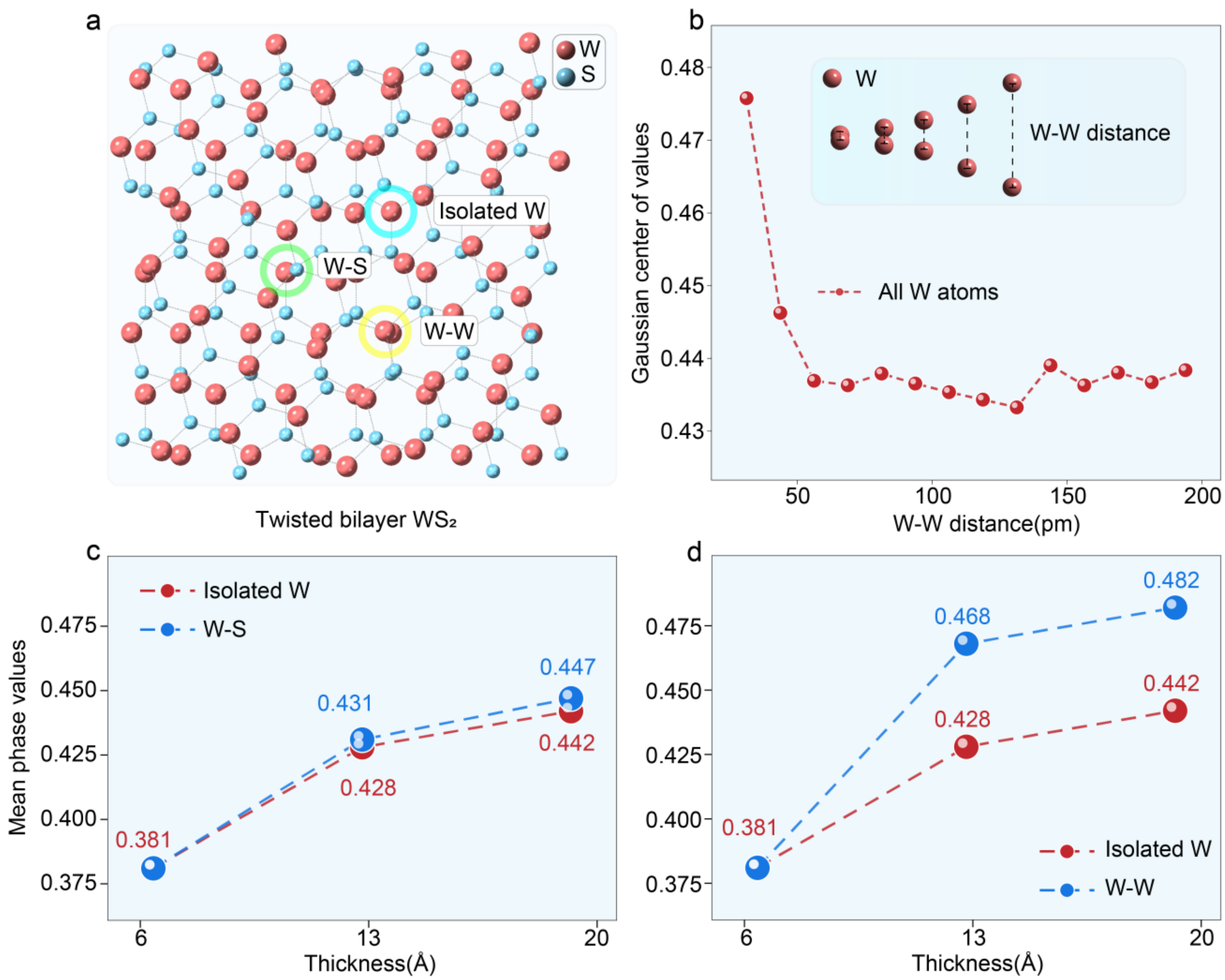

**Fig.3 | Influence of local atomic overlap on W-column phase values.** (a) Schematic illustration of three representative local projected configurations in twisted multilayer $WS_2$, including isolated W, W-S overlap and W-W overlap; red spheres denote W atoms and cyan spheres denote S atoms. (b) W-column phase-values variation as a function of projected overlap distance. (c) W-column phase values as a function of thickness for isolated W and W-S overlap. (d) W-column phase values as a function of thickness for isolated W and W-W overlap.

**Simulated thickness- and overlap-dependent phase responses**

We constructed a tt$WS_2$ simulation model to simulate the observed phase responses. The simulations were used to examine the thickness dependence and the phase enhancement associated with local atomic overlap. We analyzed the ptychographic phase response of simulated structure using the same statistical framework as for the experimental data. Like the experimental specimen, the simulated structure contains monolayer, bilayer and trilayer regions, enabling a direct comparison between experiment and simulation. In the simulation model, W atoms in the first, second and third layers are shown in red, blue and yellow, respectively, and S atoms are shown in gray (Fig. 4a). Figures 4b and 4c show the simulated W-column phase distributions and regional mean values, respectively (Supplementary Section S6). The mean phase values of the simulated monolayer, bilayer and trilayer regions are approximately 0.40, 0.45 and 0.47, respectively. This corresponds to an increase of about 12.5% from monolayer to bilayer, but only about 4.4% from bilayer to trilayer. Accordingly, the absolute increment decreases from 0.05 between the monolayer and bilayer regions to 0.02 between the bilayer and trilayer regions, reproducing the decreasing incremental response observed experimentally. To examine the simulated local-overlap effect, the W-column phase values was analyzed as a function of projected W-W distance (Fig. 4d and Supplementary Section S7). As in the experiment, the simulated phase response varies strongly at projected W-W distances below approximately 40 pm. Based on this range, the simulated phase-values distributions for different overlap configurations were compared (Figs. 4e, f and Supplementary Section S8). The simulations show that W-S overlap produces only limited phase enhancement, whereas W-W overlap produces the largest phase increase among the three configurations. This trend agrees well with the experimental W-W enhancement of about 9.3% in the bilayer region and 8.1% in the trilayer region, showing that local atomic overlap, particularly short-distance W-W projected overlap, contributes to the phase enhancement in multilayer $WS_2$. Overall, the simulations reproduce the experimentally observed dependence on layer number and projected W-W overlap. The agreement is consistent with a contribution from multiple scattering to the reconstructed phase response.

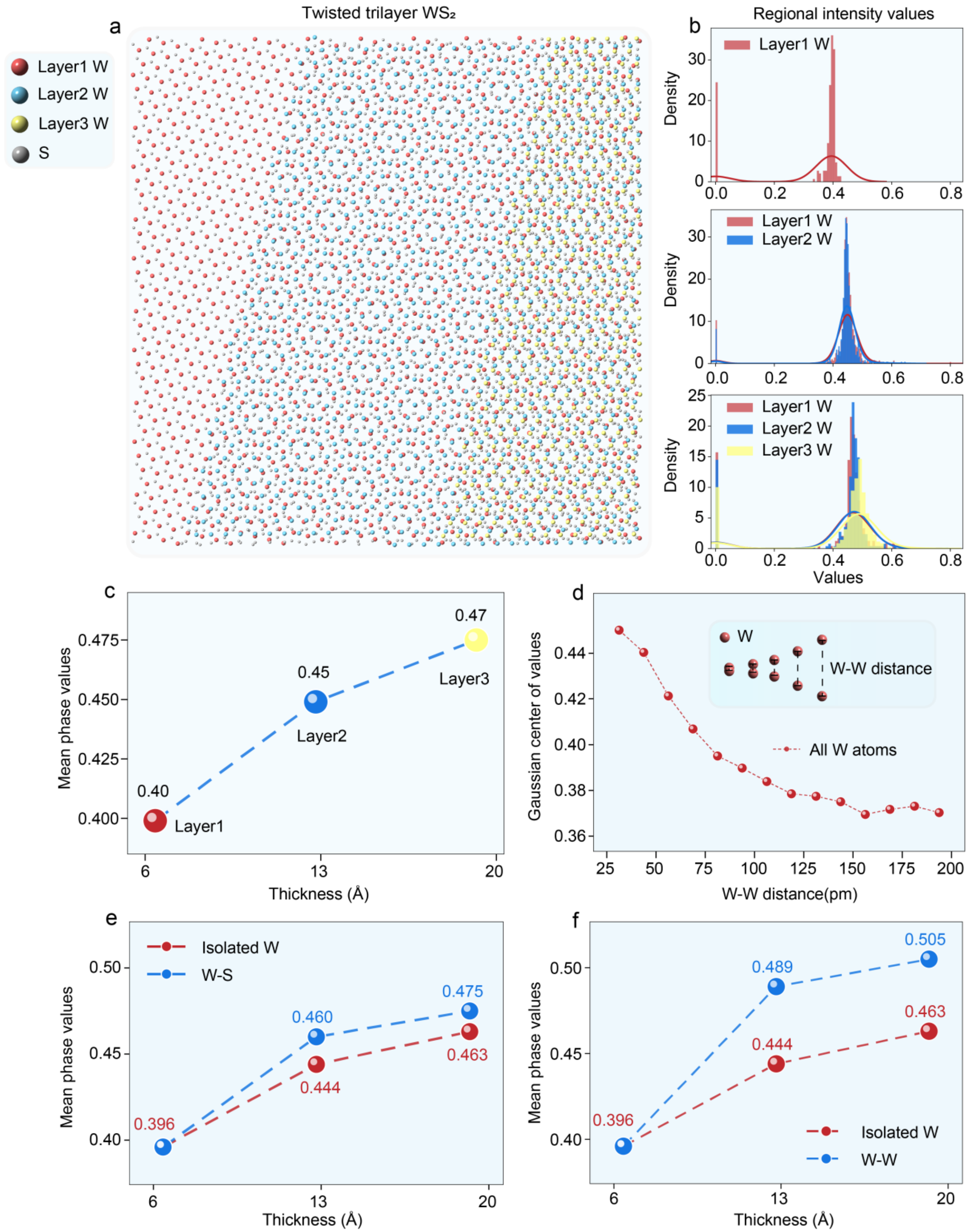

**Fig.4 | Simulations verify phase distortions induced by thickness and local overlap in $ttWS_2$.**
(a) Structural model of $ttWS_2$ used in the simulations. Red, blue and yellow W atoms correspond

to the first, second and third layers, respectively, and gray atoms denote S atoms. (b) Simulated W-column phase-values distributions in regions with different layer numbers, with red, blue and yellow corresponding to the monolayer, bilayer and trilayer regions, respectively. (c) Simulated regional mean phase values as a function of thickness. The mean phase intensities of the monolayer, bilayer and trilayer regions are approximately 0.40, 0.45 and 0.47, respectively. (d) Simulated W-column phase values as a function of projected overlap distance. (e) Simulated comparison between isolated W and W-S overlap, showing that W-S overlap causes only limited phase enhancement. (f) Simulated comparison between isolated W and W-W overlap, showing that W-W overlap produces the most pronounced phase enhancement, consistent with the experimental trend.

## Conclusion

In this work, we evaluated quantitative ptychographic phase in multilayer $WS_2$ using monolayer, bilayer and trilayer regions within a single field of view as an internal reference. The mean W-column phase increases by approximately 13.2% from the monolayer to the bilayer region and by 4.7% from the bilayer to the trilayer region. Local overlap analysis shows that W–S overlap has little effect on the W-column phase, whereas short-range projected W–W overlap produces an enhancement of approximately 8–9%. Multislice simulations reproduce both the layer-number-dependent response and the W-W overlap enhancement. Overall, these findings show that specimen thickness, multiple scattering and projected atomic overlap jointly affect the reconstructed phase. They provide a reference for for interpreting phase variations in relation to three-dimensional atomic structure and local electrostatic properties.

## Methods

### Sample preparation

The tt$WS_2$ samples were synthesized on $SiO_2$ (300 nm)/Si substrates using a custom-built two-zone furnace equipped with a 60 mm-diameter quartz tube. Two inner quartz tubes, each 1.3 mm in diameter, served as carrier-gas pathways for transporting sulfur (S) and tungsten (W) sources. The two furnace zones were heated to 930 °C and 800 °C over 40 min. Sulfur and $WO_3$ powders were transported to the substrate by Ar gas, while the pressure inside the quartz tube was maintained at approximately 0.1 kPa. The reaction was carried out for 30 min, followed by natural cooling for 15 min. The furnace was then opened to allow rapid cooling. Transmission electron microscopy samples were prepared using a PMMA-assisted wet-transfer method. After delamination and repeated rinsing in deionized water, the floating PMMA/tt$WS_2$ membrane was scooped onto Quantifoil holey-carbon TEM grids and baked on a hot plate to improve adhesion and remove residual water. Finally, the PMMA layer was removed with acetone.

**4D-STEM dataset collection**

All in-focus 4D-STEM datasets of tt$WS_2$ used in this work were acquired on a Thermo Fisher Themis Z microscope equipped with an EMPAD detector at an accelerating voltage of 60 kV. The probe-forming aperture semi-angle was set to 17.9 mrad, and the scan step size was set to 0.4219 Å to ensure sufficient probe overlap in real space and adequate sampling in reciprocal space. The camera length was set to 230 mm, corresponding to a maximum detector collection angle of approximately 97.3 mrad, which helps extend the spatial resolution of the ptychographic reconstructions. Each 4D-STEM dataset consisted of 256 × 256 diffraction patterns, and each diffraction pattern contained 128 × 128 pixels. The exposure time for each scan position was 1 ms, resulting in a total acquisition time of more than 1 min for each dataset. Owing to microscope instabilities during acquisition, the ADF images generated from the 4D-STEM datasets show noticeable affine distortions. In addition, because of the relatively large chromatic aberration of the microscope ($C_c$ = 1.5 mm) and the energy spread of the Schottky field-emission gun (1.0 eV), the effective spatial resolution of the ADF images generated from the 4D-STEM datasets is approximately 2.1 Å.

**Ptychography reconstruction**

The ptychography reconstruction was applied to the 4D-STEM datasets to improve spatial resolution and localization precision and resolve the atomic structure of tt$WS_2$. The quantitative interpretation of electron ptychography depends on accurate scan-position calibration. Before iterative reconstruction, the uniform affine transformation between the scan and camera coordinates was determined using the hybrid calibration method [37]. Taking the dataset shown in Figure 1 as an example, the calibrated scanning vectors along the fast and slow scan directions were [0.4215, 0.0175] Å and [−0.0156, 0.4184] Å, respectively, indicating noticeable compression of the STEM frame along the slow scan direction. The iterative ptychographic reconstruction was then performed using the LSQML engine. During the reconstruction, residual positional errors were corrected following the iCGD protocol [38].

### Quantitative analysis of reconstructed phase images

The analysis of reconstructed phase images comprised accurate atomic-column identification, phase-values extraction and classification of local atomic overlap. The reconstructed complex object function was first converted into a phase image, and the effective region containing monolayer, bilayer and trilayer $WS_2$ was selected for further analysis (Supplementary Section S1). The crystallographic orientations of different layers were determined from the Fourier transform of the reconstructed phase image, and the corresponding unit-cell positions were initially identified using periodic lattice signals and template matching (Supplementary Section S2). For sub-pixel localization and quantitative phase extraction, the detected unit cells were analyzed using a Gaussian Mixture Model (GMM)-based fitting method. Each atomic column was described by an individual Gaussian component. This procedure provided the optimized position, amplitude and width parameters of each atomic column, allowing the integrated phase values of each column to be quantitatively calculated from the fitted Gaussian parameters. After excluding boundary regions, contaminated areas and defective regions using manually defined masks, histograms of the integrated W-column phase were generated for the monolayer, bilayer and trilayer regions to quantify the thickness-dependent phase response (Supplementary Section S3). To evaluate the influence of local stacking, the projected distances between atomic columns were calculated, and W columns were classified into isolated W, W-S overlap and W-W overlap configurations (Supplementary Sections S4 and S5). The phase intensities of these configurations were compared to determine how specimen thickness and local atomic overlap modify the quantitative ptychographic phase.

### 4D-STEM simulations using abTEM

4D-STEM simulations were performed on a tt$WS_2$ model using the abTEM package to test the origin of the experimentally observed thickness- and overlap-dependent phase variations. The simulated structure contained monolayer, bilayer and trilayer regions for direct comparison with the experimental thickness-staircase specimen. The datasets were generated using the multislice method with parameters matched to the experiment. These comprised a 60 kV accelerating voltage, 17.9 mrad probe-forming aperture semi-angle, 0.5 Å Gaussian source FWHM and 0 Å defocus. Sampling used a 256 × 256 scan grid, 0.4219 Å scan step size, 128 × 128 detector pixels and a maximum scattering angle of 97.3 mrad. Ptychographic reconstruction and subsequent statistical analysis were performed using the same workflow as for the experimental data to compare the phase responses of monolayer, bilayer and trilayer regions, as well as isolated W, W-S overlap and W-W overlap configurations.

**Data Availability**

The data that support the findings of this study are available from the corresponding author on reasonable request.

**Additional Information**

Supporting Information. The online version contains supplementary material available at https://.

**Acknowledgements**

S.C.N. acknowledges support from the National Natural Science Foundation of China (52403303). A.M.N. acknowledges support from the National Natural Science Foundation of China (52525207 and 52288102) and the Natural Science Foundation of Hebei Province (E2024203054). T.X. acknowledges financial support from the Scientific Research Innovation Capability Support Project for Young Faculty (SRICSPYF-BS2025048).

**Author Contributions**

A.M.N. and S.C.N. supervised the project. K.Y., T.X. and Z.Y.L. synthesized the $WS_2$ samples. S.C.N., Y.H.Z and Q.L.F. performed the 4D-STEM characterization. S.C.N., H.B. and J.T. analyzed the data. H.B., S.C.N., J.T. and A.M.N. wrote the manuscript with input from all authors. All authors reviewed the manuscript before submission.

**Competing Interests**

The authors declare no competing interests.